\documentclass[a4paper]{article}
\usepackage{amstext}
\usepackage[dvips]{graphics}
\begin{document}
% \draft
\title{Measurements of the Solar Neutrino Flux from Super-Kamiokande's 
First 300 Days}
\author{ The Super-Kamiokande Collaboration}
\date{\it Submitted to Phys. Rev. Lett.}
\maketitle

\begin{center}
%ICRR
Y.Fukuda$^a$, T.Hayakawa$^a$, E.Ichihara$^a$, K.Inoue$^a$,
K.Ishihara$^a$, H.Ishino$^a$, Y.Itow$^a$,
T.Kajita$^a$, J.Kameda$^a$, S.Kasuga$^a$
% ...

\footnotesize \it

$^a$Institute for Cosmic Ray Research, University of Tokyo, Tanashi,
Tokyo 188-8502, Japan\\
% ...

\end{center}

\begin{abstract}

   The first results of the solar neutrino flux measurement 
from Super-Kamiokande are presented.
   The results shown here are obtained from data taken 
between the 31st of May, 1996, and the 23rd of June, 1997. 
%15th of June, 1997.
Using our measurement of recoil electrons with energies above 6.5 MeV,
we infer the total flux of $^{8}$B solar neutrinos to be
2.42$\pm$0.06(stat.)$^{+0.10}_{-0.07}$(syst.)$\times$10$^6$/cm$^2$/s.
This result is consistent with the Kamiokande measurement and is 36\% of the 
flux predicted by the BP95 solar model.
   The flux is also measured in 1.5 month subsets and shown to be consistent
with a constant rate.
\end{abstract}
%%\pacs{96.60.Kx,95.85.Qx,96.40.Tv}

%body of paper here
%%\narrowtext

%   The Sun produces neutrinos in its central core through 
%nuclear reaction chains, and these 
%solar neutrinos play crucial roles both in astrophysics and in particle 
%physics. 

The neutrino plays a crucial role in both astrophysics and particle
physics. This report is on measurements of solar neutrinos that are
 produced in the core of the sun through nuclear reaction
chains.  Since neutrinos pass through matter largely unimpeded, the
mechanism of solar energy generation taking place at the central core
of the sun can be studied directly by solar neutrinos.  Evidence of as
yet unresolved neutrino properties may also be obtained by detailed
studies of the solar neutrinos, as these neutrinos are produced in
very dense matter, pass through the core and surrounding layers to the
surface of the star, and reach the Earth after about 150 million
kilometers of flight.  This naturally-arranged situation enables us to
study possible neutrino mass and magnetic properties.

%   Several tens of billions of those neutrinos traversing each square cm of 
%the Earth every second have been detected by four different
%experiments

Tens of billions of neutrinos from the sun traverse each square
centimeter of the Earth every second. Four different experiments
\cite{Cl37,KMII,KMIII-PRL,Sage,Gallex} have detected these neutrinos.
   All of the experiments have observed a significantly lower flux than 
that predicted by 
standard solar models (SSMs)\cite{SSM-BP95,SSM-BP92,SSM-TCL}.
   This discrepancy, first suggested by the historic
Cl-experiment\cite{Cl37}, is called ``the solar neutrino problem."
   One of the characteristics of the problem is that the amount of 
suppression among the experiments appears to be energy-dependent.
   Detailed studies of this phenomenon 
strongly suggest that these deficits are 
not easily explained by changing 
solar models, but can be naturally explained by neutrino 
oscillations\cite{Hata}.

%   It is, however, strongly desirable to observe direct and 
%solar-model-independent evidence for neutrino oscillations.
%   Super-Kamiokande, which is able to measure
%a distortion of the recoil electron energy spectrum,   
%short-term time variations (like the daytime and nighttime flux difference),
%and the time variation associated with the solar activity cycle, has such an 
%ability.

Super-Kamiokande is capable of observing direct and
solar-model-independent evidence of neutrino oscillations. Observation
of a distortion of the recoil electron energy spectrum, short-term
time variations (like the daytime and nighttime flux difference), and
the time variation associated with the solar activity cycle, would
constitute such evidence.  The production of solar neutrinos is
supposed to be stable over a time scale of several million years, and
if any time variations were to be found, that would indicate either neutrino
mass and mixing\cite{MSW}, or the presence of non-zero neutrino
magnetic moments\cite{MagM}.  Such solar-model-independent studies
require small statistical and systematic errors, which in turn
require long-term data accumulation.

   In this letter, we present the first results of the flux measurement 
from Super-Kamiokande. 
The daytime and nighttime fluxes and the flux for each 
1.5 month interval over one year of data-taking are also discussed.

   Super-Kamiokande, the first ``second-generation'' solar neutrino experiment,
started operation on April 1st, 1996.  Located at a depth of 
2700 meters water equivalent in the Kamioka Mozumi mine in Japan, the detector
is a 50,000 ton imaging water Cherenkov detector and has 
a cylindrical geometry, 39.3 m in diameter and 41.4 m in height. 
   The central 32,000 tons -- 36.2 m in height and 33.8 m in diameter -- is 
called the inner detector and is viewed by 11,146 50-cm 
photomultiplier tubes (PMTs) which cover 40\% of the inner surface.
   Surrounding the inner detector is the outer detector, which comprises 
a 2.6 to 2.75~m 
thick layer of water.
The outermost 2.15 m of this water is an active 
detector, viewed by 1,885 20-cm PMTs to 
identify in-coming particles.
This also serves to passively reduce $\gamma$ and neutron backgrounds from the rocks
surrounding the detector.
   An inactive region of 0.6 m thickness separates the inner
detector from the active part of the outer detector with black sheets
which prevent light transmission between the two regions.
   A fiducial volume of 22,500 tons of water, about 70\% of the inner volume, 
is used for the solar neutrino analysis; 
   the outer edge of this fiducial volume is located 2 m from the surface of
the inner detector.
%   There is, therefore, a water thickness of 
This gives 4.75 m of water outside of the fiducial 
volume -- 13.2 radiation lengths and 7.9 nuclear collision lengths -- a very 
thick shield against backgrounds from the rock.

Solar neutrinos produce electrons in the water through
neutrino--electron elastic scattering, and subsequently the recoil
electrons emit Cherenkov photons.  These photons are then detected by
the PMTs on the surface of the inner detector.  The front-end
electronics for each hit PMT creates a 200 ns wide pulse, and the
hardware trigger is made via a simple sum of the number of hit PMT
pulses.  The trigger threshold for events used in this analysis, 29
hits within an approximately 200 ns coincidence window in excess of 
the continuous noise hits due to PMT dark noise,
corresponds to about 5.7 MeV (total energy).  
The analysis threshold was set at 6.5 MeV, for which the hardware trigger 
is only 0.2\% inefficient. 
The trigger rate during this period was stable at $\sim$11Hz.

Solar neutrino interactions are reconstructed by using the charge and
timing data from the hit PMTs.  Recoil electrons have a short range
($< 8$ cm) and their Cherenkov light effectively comes from a point.
We use a grid search method to find the event vertex.  First, hits to
be used for the vertex search are selected by sliding a time window
until maximum signal to background significance is obtained.  Then the
grid point which gives the best fit on a 4m fixed mesh is
selected.  Final refinement of the vertex position, down to a 6 cm
step size, is also done by a grid search method.  The direction of
each event is obtained by a maximum likelihood technique which uses
the relative direction of each hit PMT.  For each event, the recoil
electron energy is initially determined by the number of hit
PMTs in a 50 ns time window, since most of the PMT hits for events in
the energy range of interest are due to only one photoelectron.   
   We make corrections to the number of hit PMTs in order to compensate for
light attenuation through the water, bad PMTs, angular dependence of
the acceptance, the effective density of PMTs, and the probability of
a two photon hit on a PMT in order to get uniform responce over the fiducial
volume.  We further correct for noise hits due to
the PMT dark rate ($\sim$3.3kHz) which contributes about 1.8 hits
within 50 ns.  The tail of the time distribution up to 100 ns, caused
by scattering of light in the water and reflections on the surfaces of
the PMTs and black sheet, is also corrected for.  The resulting
corrected number of hit PMTs, $N_{eff}$, is closely related to the
energy of the events.

   An electron linear accelerator (LINAC) is used for calibrating the
absolute energy scale, angular resolution, and vertex position resolution.
   Details of the LINAC calibration will be described elsewhere; a brief 
summary of the calibration is given here.
   The LINAC, located near the Super-Kamiokande detector,
injects mono-energetic electrons with a tunable energy ranging from 5 MeV to
16 MeV.
   This matches the energy of the solar neutrinos detected in Super-Kamiokande.
%   The beam is transported into the detector through a 10 cm diameter beam 
%pipe, while
%   the beam energy is determined by a system of bending magnets and 
%collimators.
%   The energy spread is less than 0.3\%.
%   At the exit of the beam pipe is a titanium window of 100 $\mu$m thickness,
%   and a 1 mm thin plastic scintillator is mounted before the titanium window 
%to produce a trigger signal for LINAC electrons.
   The absolute energy of the beam is measured by a germanium detector, 
which was in turn calibrated by gamma-ray sources and internal-conversion electrons from a $^{207}$Bi source;
the uncertainty of the beam energy is less than 20~keV over the energy range
covered by the LINAC.
    LINAC data were taken at 6 different positions in the 
Super-Kamiokande tank between December 1996 and October 1997\cite{linac}.

\begin{figure}[h]
\center
\resizebox{12cm}{!}{\includegraphics{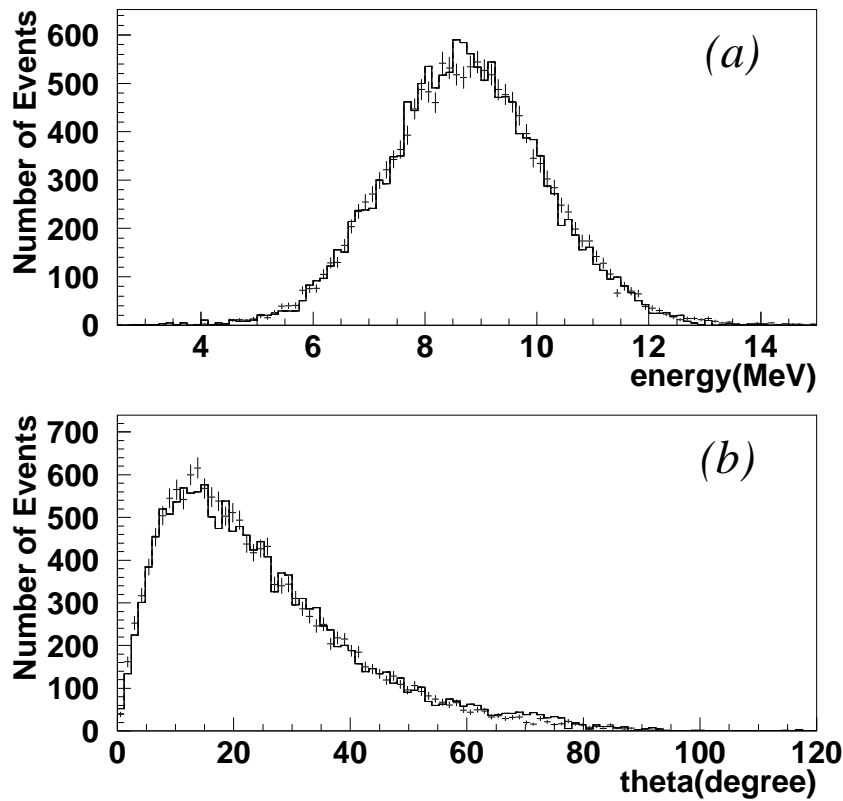}}
\caption{
(a) A typical energy distribution of 8.86MeV electrons produced by 
the electron LINAC. 
(b) A typical angular distribution of 8.86MeV electrons produced by 
the electron LINAC. 
Also shown is that of the Monte Carlo electron events produced at the same
vertex position.
}
\label{fig:LINAC-ang}
\end{figure}

   A typical distribution of reconstructed electron energy and
direction relative to the injected beam direction is 
shown in Fig.\ref{fig:LINAC-ang} for an 8.86 MeV/c momentum beam together 
with that obtained by a Monte Carlo simulation.
   Parameters in the Monte Carlo program,  mainly the photon scattering and 
absorption lengths in water, were tuned in such a way that the Monte
Carlo  reproduces the LINAC data at various positions and energies.
The Monte Carlo calculations are used to extrapolate these calibrations 
to the entire fiducial volume and all directions.
%%The estimated energy, angular, and vertex resolutions for 10 MeV electrons
%%are ???, ???, and, ???, respectively.

   The energy calibration by the LINAC is cross-checked using
gamma-rays from Ni(n,$\gamma$)Ni reactions.
   The absolute energy scale obtained by the Ni(n,$\gamma$)Ni is
produced by comparing the observed gamma-ray spectrum with a simulation
based on a model of gamma transition; it is 1.4\% lower than
that obtained by the LINAC calibration.
   It is suspected that the difference is due to gamma-rays with
small branching ratios in the Ni(n,$\gamma$)Ni reaction and
non-uniformity of the source. 
   Hence, we use the Ni(n,$\gamma$)Ni calibration only for determining the 
relative position dependence of the energy scale in the detector.
   The comparison of the position dependence between the LINAC and the
Ni(n,$\gamma$)Ni shows that the agreement is better than 0.5\%.
   The absolute energy scale is also cross-checked using beta-decays of
$^{16}$N produced by cosmic ray stopping muons.
The observed beta spectrum agrees with the Monte Carlo 
simulation to better than 1.5\% within the statistical accuracy of 
the measurement.

Water transparency used as inputs for a Monte Carlo program was measured by 
a dye laser and a CCD camera.
   Since the water transparency varies slightly with time, we monitor and 
correct for the small changes by using the Michel spectrum of electrons 
originating from stopping muons. 
   We observe about 1,200 such events per day.
   Using this correction, we kept the peak of the Michel spectrum 
stable to within $\pm$0.5\%.
   In addition, the time dependence of the peak energy of the 
 $\gamma$-rays from the Ni-calibration after this correction is less than
$\pm$0.5\%.
   A similar check was also made by using muon induced spallation events
and a similar result was obtained.

   For the present analysis we have used the data obtained from 297.4 live days
between 31 May 1996 and 23 June 1997.  
The detector live time fraction during this period was greater than 90\%.
   Most of the down time was due to calibrations of the detector.
  
The data set consisting of $\sim$3$\times$10$^8$ events was reduced
using algorithms similar 
to those used in the Kamiokande
experiment~\cite{KMII,KMIII-PRL}. This reduction process
required that events be (1) contained, (2) low energy, and (3)
separated by more than 20 $\mu$s from any previous trigger. The
contained event cut required less than 20 hits in the outer detector
and the low energy cut required that the event have less than 1000
photoelectrons (110$\sim$120 MeV).
   We applied noise cuts that are effective in removing backgrounds in the lower 
energy region below 7 MeV.
   One of the backgrounds in this region involves events which have
small clustered hits, i.e., several hit PMTs located next to each other. 
   It is suspected that these events may come from radioactive 
contamination in the PMT glass.
   These noise cuts reduced the background level by more than a factor of three
between 6.5 MeV and 7 MeV while keeping more than 90\% of the 
signal events in the same energy region. 
   The efficiency for these cuts was obtained by Monte Carlo calculation
and also by studying the effect on spallation events. 
%   The details of these and other cuts will be explained in 
%a longer paper\cite{full-p}.
   An energy cut of 6.5 MeV yielded about 
57 events/day/kton with an overall efficiency of 94.2\%
   
   The main source of the remaining background events is muon--induced
spallation products. These decay products of fragmented $^{16}$O nuclei
can effectively mimic solar neutrino events.
We identify them via a likelihood analysis on the variables: 
time from previous muon event; distance from previous muon track; 
and muon energy loss along that track.
A cut on the likelihood function was chosen
to optimize the effectiveness of the overall spallation cut, thereby 
yielding approximately the maximum significance 
for the solar neutrino signal/$\sqrt{\rm background}$.
This reduced the event rate to 12 events/day/kton.
Application of the spallation cut results in a dead time and a 
dead volume, which we treated in analysis as an effective dead 
time for the solar neutrino signal.
   This dead time was calculated to be 20\% 
by using real muon data and 
distributing Monte Carlo low energy events randomly in 
space and time throughout the detector volume.
The data was futher reduced by removing the gamma-ray backgrounds from
the rocks surrounding the detector, 
thus giving a final rate of 7.6 events/day/kton\cite{KMII,KMIII-PRL}.

\begin{figure}
\center
\resizebox{12cm}{!}{\includegraphics{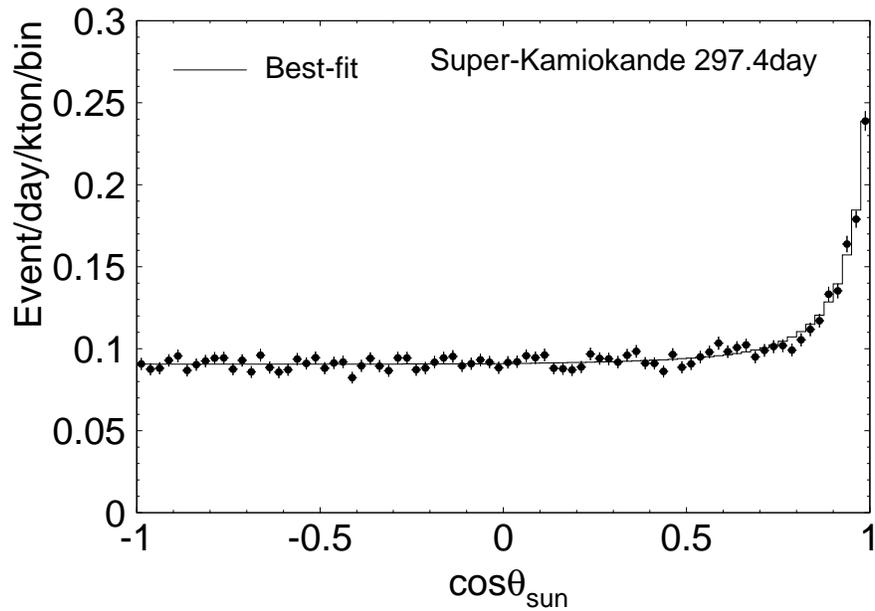}}
\caption{
  Plot of the cosine of the angle between the electron direction and a radius 
vector from the Sun. 
  One obtaines a clear peak from the solar neutrinos.
  The solid line shows the best fit to the data.}
\label{fig:cos-theta}
\end{figure}

   The directional distribution to the Sun of events in the final data sample
is shown in Fig.~\ref{fig:cos-theta}.
   The data was divided into 16 energy bins. 
The number of solar neutrino events was extracted from the binned data
by a maximum likelihood 
method using angular distributions expected for the solar neutrino
signal and 
%an estimated background distribution which is consistent
%with being flat
a near-flat background distribution with small corrections made for a slight
directional anisotropies in local detector coordinates
\cite{KMII,KMIII-PRL}.
   In this method, the $^8$B solar neutrino spectral shape is assumed.
   We obtained
4017$\pm$105(stat.)$^{+161}_{-116}$(syst.)
solar neutrino events between 6.5 MeV and 20 MeV.
   Using this number and assuming the $^8$B solar neutrino energy 
spectrum\cite{Rad-corr},
the total $^8$B solar neutrino flux was 
calculated to be
2.42$\pm$0.06(stat.)$^{+0.10}_{-0.07}$(syst.)$\times$10$^6$/cm$^2$/s,
which is consistent with the Kamiokande flux of 2.80$\pm$0.19(stat.)
$\pm$0.33(syst.)$\times$10$^{6}$/cm$^{2}$/s.
   Using this flux measurement and the most recent SSM (BP95) 
calculation \cite{SSM-BP95}, which has gone up by 16\% from 
BP92 \cite{SSM-BP92}, we get a Data/SSM of 
0.358$^{+0.009}_{-0.008}$(stat.)$^{+0.014}_{-0.010}$(syst.) 
for the 6.5MeV energy threshold data sample.

   The largest of our systematic errors comes from the uncertainty of the 
angular resolution.
   The total systematic errors of $^{+4.0}_{-2.9}$\% include 
the uncertainty of the energy determination $^{+2.3}_{-2.1}$\%, 
the uncertainty of the expected $^8$B energy spectrum $^{+1.2}_{-1.1}$\%, 
the uncertainty in trigger efficiency (+0.2\%), 
the noise cuts ($\pm$0.7\%), directional fit (+2.9\%), 
data reduction ($\pm$0.2\%), background shape($\pm$0.1\%), 
spallation dead time ($<$0.1\%), 
fiducial volume (-1.3\%), cross section ($\pm$0.5\%), and live time 
calculation ($\pm$0.1\%). 
 The details of the systematic errors will be explained in \cite{full-p}.

   Solar neutrino fluxes for different data sets were obtained. 
   The number of events extracted above 7 MeV was 
3362$^{+96}_{-88}$(stat.)$^{+138}_{-101}$(syst.) and the total flux from this
data set is
2.44$^{+0.07}_{-0.06}$(stat.)$^{+0.10}_{-0.07}$(syst.)$\times$10$^6$/cm$^2$/s.
   The flux for the smaller inner fiducial volume of 11.7 kton was 
2.47$^{+0.09}_{-0.08}$(stat.)$^{+0.10}_{-0.07}$(syst.)$\times$10$^6$/cm$^2$/s.
   Both are  consistent with that obtained for the entire 22.5 kton fiducial 
volume above 6.5 MeV.
   The daytime and nighttime fluxes were measured separately. 
The daytime flux was
2.39$\pm$0.09(stat.)$^{+0.10}_{-0.07}$(syst.)$\times$10$^6$/cm$^2$/s
and the nighttime flux was 
2.44$^{+0.09}_{-0.08}$(stat.)$^{+0.10}_{-0.07}$(syst.)$\times$10$^6$/cm$^2$/s.
   There is no significant difference seen between the daytime and nighttime fluxes. These fluxes are interesting for
studying neutrino oscillations\cite{Hata}, and a detailed study and  
consideration of the implications of the day/night results will be published 
later.

\begin{figure}
\center
\resizebox{12cm}{!}{\includegraphics{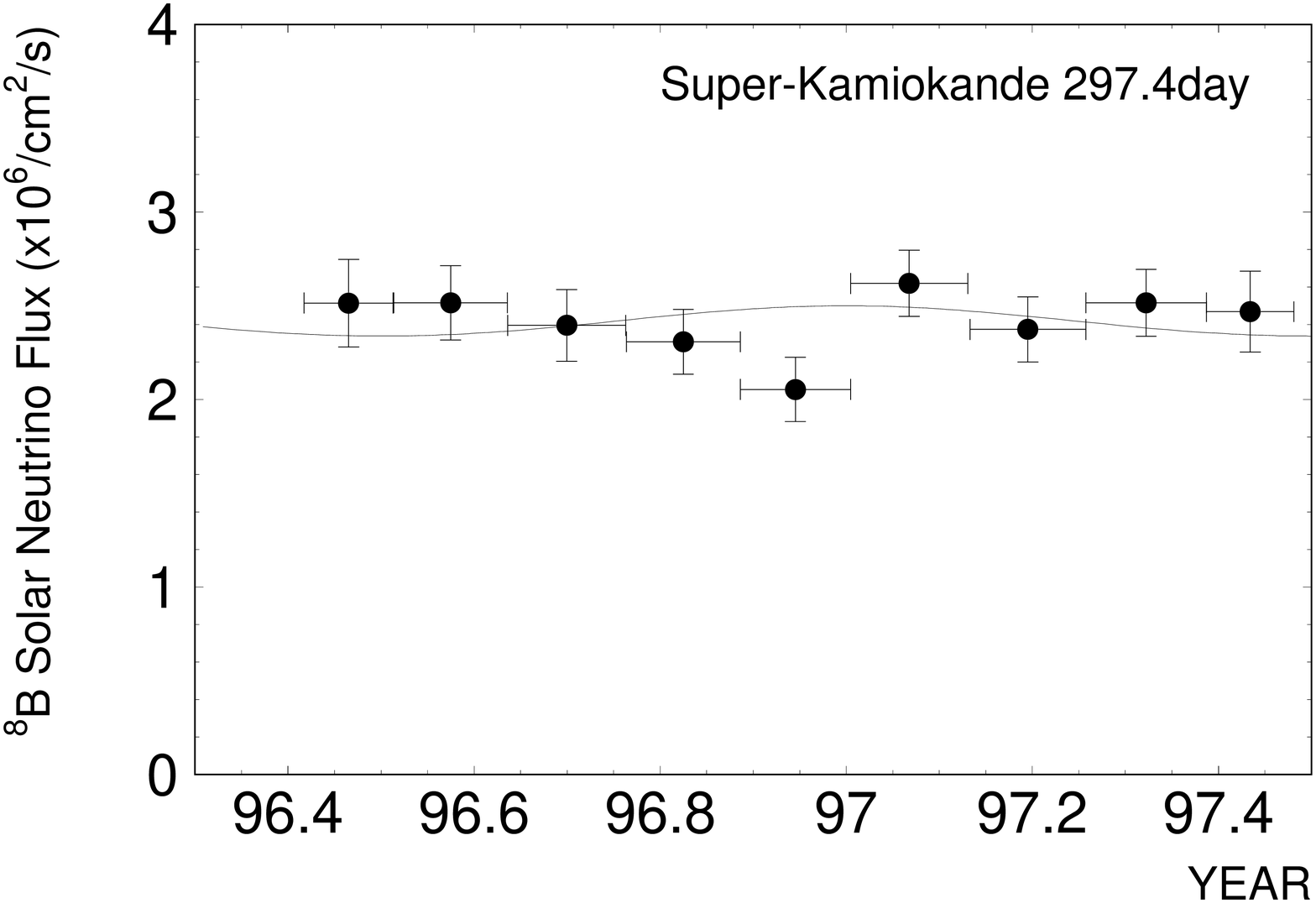}}
\caption{
The flux in $\sim$1.5 month periods from June 96 to June 97.
The solid line shows the expected yearly change of the flux (about 7\%) due to
the eccentricity of the Earth's orbit around the Sun.}
\label{fig:seasonal}
\end{figure}

        Finally, the data were divided into subsets, each 
consisting of about 1.5 months 
of data where those divisions were determined by taking into account the 
date of Earth's perihelion (Jan-2) and aphelion (Jul-3) in its orbit around the Sun. 
   In this way we are able to
study the stability of the solar neutrino flux and look for possible 
seasonal effects, although the statistics have not yet reached the
level needed to study the effect of the 'Just So' oscillation 
scenario\cite{JustSo,Hata}. 
   Fig.~\ref{fig:seasonal} shows that the flux 
was stable over a one year long period. The anticipated flux
variation ($\sim$7\% maximum) due to the eccentricity of the Earth's 
orbit is shown by the solid line. The $\chi^{2}$ for the solid line is 10.30 
with 8 degrees of freedom.

   We also performed an independent solar neutrino 
analysis \cite{Zthesis} in addition to the Kamiokande-based analysis described here \cite{Jthesis}. 
   The two analyses had access to the same raw calibration data and both used 
the same raw data. 
   All subsequent steps in the data processing, event reconstruction, and 
efficiency determination for the two analyses were different and were 
performed independently. The results from that analysis are in
agreement with the flux measurement reported here. A detailed comparison will
be presented in a future publication. 

In summary, the Super-Kamiokande detector has observed a stable flux of solar
neutrinos that is consistent with that reported by the Kamiokande
experiment and is significantly lower than predicted by standard solar
models.
    The implications of the day/night flux ratio and the spectrum of 
the recoil electron energy, both very important for the study of a possible
neutrino mass, will be published later.
\\
\\
   We gratefully acknowledge the cooperation of the Kamioka Mining and Smelting
Company. 
   This work was partly supported by the Japanese Ministry of Education,
Science and Culture and the U.S. Department of Energy.

%
% now the references. delete or change fake bibitem. delete next three
%   lines and directly read in your .bbl file if you use bibtex.
%%\begin{references}

%%%%%%%%%%%%%%%%%%%%%%%%%%%%%%%%%%%%%%%%%%%%%%%%%%%%%%%%
%%%%%%%%%%%%%%%%%%%%%%%%%%%%%%%%%%%%%%%%%%%%%%%%%%%%%%%
%
%%%%%%%%%%%%%%%%%%%%%%%%%%%%%%%%%%%%%%%%%%%%%%%%%%%%%%
%
%%%%%%%%%%%%%%%%%%%%%%%%%%%%%%%%%%%%%%%%%%%%%%%%%%%%%


\begin{thebibliography}{99}
\bibitem{Cl37} B.T.Cleveland et al., Nucl. Phys. B(Proc. Suppl.)
               38, 47(1995); R.Davis, Prog. Part. Nucl. Phys. 32, 13(1994). 
\bibitem{KMII} K.S.Hirata et al., Phys. Rev. Lett. 65, 1297(1990);
               K.S.Hirata et al., Phys. Rev. D44, 2241(1991); D45, 2170E(1992).
\bibitem{KMIII-PRL}Y.Fukuda et al., Phys. Rev. Lett. 77,1683(1996).
\bibitem{Sage} J.N.Abdurashitov et al., Phys. Lett. B328, 234(1994).
\bibitem{Gallex} P.Anselmann et al., Phys. Lett. B327, 377(1994):
                 B342, 440(1995).
\bibitem{SSM-BP95}J.N.Bahcall and M.Pinsonneault, ReV. Mod. Phys. 67, 
781(1995).
\bibitem{SSM-BP92}J.N.Bahcall and M.Pinsonneault, ReV. Mod. Phys. 64, 
885(1992).
\bibitem{SSM-TCL} S.Turck-Chi$\grave{\text{e}}$ze and I.Lopes, Ap. J. 408,347(1993).
\bibitem{Hata} See for example, N. Hata and P. Langacker, IASSNS-AST 97/29
               for a recent review.
\bibitem{MSW} S.P.Mikheyev and A.Y.Smirnov, Sov. Jour. Nucl. Phys. 
              42, 913(1985);
              L.Wolfenstein, Phys. Rev. D17, 2369(1978).
\bibitem{MagM} L.B.Okun et al., Sov. J. Nucl. Phys. 44,440(1986);
               C.S.Lim and W.J.Marciano, Phys. Rev. D37,1368(1988);
               E.Kh.Akhmedov, Phys. Lett. B213, 64(1988).
\bibitem{linac} Super-Kamiokande Collaboration, to be submitted.
\bibitem{full-p} Super-Kamiokande Collaboration, to be submitted.
\bibitem{Rad-corr} J.N.Bahcall et al., Phys. Rev, D51, 6146(1995).
\bibitem{JustSo} S.M.Bilenky and B.Pontecorvo, Phys. Rep. 41, 225(1978);
                 V.Barger, R.J.N.Phillips and K. Whisnant, Phys. Rev. D24, 538 
                 (1981); 
                 S.L.Glashow and L.Krauss, Phys. Lett. B190, 199(1987).
\bibitem{Zthesis} Z.~Conner, Ph.D. Thesis, University of Maryland (1997).
\bibitem{Jthesis} Y.~Koshio, Ph.D. Thesis, University of Tokyo (1998);
 H.~Okazawa, Ph.D. Thesis, Niigata University (1998);
 T.~Yamaguchi, Ph.D. Thesis, Osaka University (1998);
 R. Sanford, Ph.D. Thesis, Louisiana State University (1998).

%\end{references}
\end{thebibliography}
\end{document}